\documentclass[twocolumn]{aastex702}

\usepackage{graphicx}
\usepackage{amsmath}

\graphicspath{{./}{figures/}}

\journalinfo{}
\makeatletter
\let\frontmatter@title@above\relax
\makeatother
\begin{document}
\title{A Failed Filament Eruption and Associated Active Region Heating Driven by External Magnetic Reconnection}

\correspondingauthor{Yuandeng Shen}

\author{Yudi Ou}
\email{ouyudi@hit.edu.cn}
\affiliation{State Key Laboratory of Solar Activity and Space Weather, School of Aerospace, Harbin Institute of Technology, Shenzhen 518055, People's Republic of China}

\author{Yuandeng Shen}
\email{ydshen@hit.edu.cn}
\affiliation{State Key Laboratory of Solar Activity and Space Weather, School of Aerospace, Harbin Institute of Technology, Shenzhen 518055, People's Republic of China}

\author{Dongxu Liu}
\email{25B961002@stu.hit.edu.cn}
\affiliation{State Key Laboratory of Solar Activity and Space Weather, School of Aerospace, Harbin Institute of Technology, Shenzhen 518055, People's Republic of China}

\author{Reetika Tiwari}
\email{reetikatiwari4@gmail.com}
\affiliation{State Key Laboratory of Solar Activity and Space Weather, School of Aerospace, Harbin Institute of Technology, Shenzhen 518055, People's Republic of China}

\begin{abstract}
We investigate a failed filament eruption and associated coronal heating in NOAA Active Region 12353 on 2015 May 23, focusing on the relationship between magnetic confinement and external reconnection. Coordinated observations from the Goode Solar Telescope and the Solar Dynamics Observatory, together with nonlinear force-free field extrapolations, allow us to trace the event from the merging of two low-lying filamentary structures to the failed eruption of the resulting elongated filament. 
Before the rapid rise, the modeled flux rope had entered a regime susceptible to kink instability, while low-atmospheric internal reconnection had already occurred; both may have contributed to the initial acceleration.
The rising flux rope remained below the torus instability threshold and was confined by a strong, asymmetric overlying field that compressed it against the overlying loops.
An elongated interface brightening, bidirectional plasmoid-like motions, newly brightened loop systems, and progressive weakening of the filament are consistent with external reconnection that changed the magnetic connectivity, removed magnetic flux from the rope, and disrupted its coherent structure. Differential emission measure analysis indicates that the external reconnection heated the active region loop system. A transient, localized structure with high-temperature emission was observed at the interaction interface, consistent with the expected location of a current sheet associated with external reconnection. These results suggest that the strong confining field halted the filament eruption and sustained the interface compression required for current sheet formation, enabling external reconnection to disrupt the flux rope and heat the active region loop system.
\end{abstract}

\section{Introduction}\label{sec:introduction}
Solar eruptions involve the rapid restructuring of coronal magnetic fields and the release of stored magnetic energy. Their major manifestations, including flares, jets, filament eruptions, and coronal mass ejections (CMEs), represent different aspects of this energy release process \citep{2001ApJ...559..452Z, 2004NewA....9..611L, 2021RSPSA.47700217S}. Solar filaments consist of cool, dense plasma supported in magnetic dips within highly sheared arcades or magnetic flux ropes above photospheric polarity inversion lines (PILs) \citep{1998SoPh..182..107M,2015ApJ...814L..17S}. Their formation and eruption therefore provide important diagnostics of how coronal magnetic energy and helicity are stored, transported, and released, as well as valuable constraints on models used for space weather forecasting \citep{2011LRSP....8....1C, 2011LRSP....8....6S}.

Filament eruptions are commonly classified as full, partial, or failed according to the fate of both the cool filament material and its supporting magnetic structure \citep{2000ApJ...537..503G,2001ApJ...549.1221G,2007SoPh..245..287G,2009A&A...498..295T}. In a full eruption, most of the magnetic structure supporting the filament escapes, usually together with a CME. In a failed eruption, the filament rises but remains confined, and most of the lifted material subsequently drains back. A partial eruption occurs when the erupting structure separates into escaping and surviving components, often through reconnection within or around the flux rope \citep{2006ApJ...637L..65G,2012ApJ...750...12S, 2012ApJ...756...59L,2021ApJ...923..142C,2022ApJ...929...85D}. Because the cool plasma and the supporting magnetic field do not necessarily have the same fate, the terms partial and failed can describe different aspects of the same event. It is therefore important to specify whether an eruption classification refers primarily to the filament material, the magnetic structure, or the associated CME.

The initiation and evolution of filament eruptions are governed by the interplay among magnetic reconnection, flux-rope instability, and confinement by the surrounding magnetic field. Internal reconnection in the low atmosphere can modify magnetic connectivity, facilitating the formation of a magnetic flux rope or contributing to the initial rise of a preexisting one \citep{2001ApJ...552..833M, 2021NatAs...5.1126J, 2024ApJ...964..125S}. As the flux rope forms and evolves, its susceptibility to the kink instability depends critically on its magnetic twist, which is commonly quantified by the twist number $T_w$ along magnetic field lines \citep{2006JPhA...39.8321B}. A sufficiently twisted flux rope may become kink-unstable and undergo axis writhing and rotation during the early eruption phase \citep{1979SoPh...64..303H,2004A&A...413L..27T,2012ApJ...756...59L}. The critical twist for triggering the kink instability depends on the magnetic configuration and typically lies within $|T_w|\sim1.25$--$2.5$ \citep{2001A&A...367..321B, 2003ApJ...589L.105F, 2003A&A...406.1043T, 2004A&A...413L..27T, 2005ApJ...630L..97T, 2011ApJ...735L..43S}.

However, reaching the kink-instability threshold and beginning to rise do not necessarily lead to a successful eruption
\citep{2003ApJ...595L.135J,2005ApJ...630L..97T}.
The subsequent evolution depends strongly on the confinement imposed by the external magnetic field overlying the flux rope.
If this external field decreases rapidly enough with height, its confining force weakens, allowing the torus instability to develop.
The rate at which the external field decreases with height is quantified by the decay index,
$n=-d\ln B_{\rm ex}/d\ln h$,
where $B_{\rm ex}$ is the strength of the external confining field and $h$ is the height.
The torus instability is typically triggered when the decay index reaches a critical value, with $n_{\rm crit}$ generally lying in the range $1$--$2$
\citep{2006PhRvL..96y5002K,2007ApJ...668.1232F,2010ApJ...708..314A,2010ApJ...718.1388D,2015ApJ...814..126Z}.
A strong overlying field that decreases slowly with height can instead keep the flux rope torus-stable.
The overlying field may then halt the rise of the flux rope, resulting in a failed eruption in which the rising magnetic structure remains confined in the low corona
\citep{2008ApJ...679L.151L,2013ApJ...778...70C,2015ApJ...804L..28S,2019ApJ...871..229C}.

Beyond the height dependence of the overlying field, its absolute strength in the low corona and the asymmetry of the overlying arcade can also contribute to the confinement of a rising flux rope and favor a failed eruption
\citep{2009ApJ...696L..70L,2015ApJ...804L..28S,2018Natur.554..211A}. Even under similar external magnetic conditions, eruption outcomes can differ. Among six successive eruptions of the same filament, five relatively slow events associated with weaker flares failed, whereas the fastest event, accompanied by the strongest flare, produced a CME. The overlying field strength, decay index, and asymmetry changed little during the sequence \citep{2011RAA....11..594S}. This comparison suggests that the available driving energy and the resulting acceleration are also important. 
An eruption may even fail after reaching a height favorable for torus instability, indicating that favorable torus conditions alone do not guarantee a successful eruption. Additional effects, including strong flux-rope rotation \citep{2019ApJ...877L..28Z}, non-axisymmetric Lorentz forces associated with the three-dimensional geometry of the flux rope \citep{2021NatCo..12.2734Z}, and reconnection with the surrounding field \citep{2023MNRAS.525.5857J}, may still alter the eruption outcome. 
Reconnection with neighboring magnetic structures may also transfer twist away from the filament-supporting magnetic structure and reduce its nonpotential magnetic energy, thereby weakening the eruption driver and contributing to a failed eruption \citep{2021ApJ...923...45Z}.
The decay index must be considered together with the available free magnetic energy, the three-dimensional geometry of the confining field, and the reconnection processes during the eruption.

External reconnection involving an erupting flux rope and the surrounding magnetic field can either facilitate or suppress an eruption, depending on which flux systems reconnect and how magnetic flux is redistributed. In the magnetic breakout model, reconnection above the erupting core transfers overlying flux away from the central arcade, thereby reducing confinement \citep{1998ApJ...502L.181A, 1999ApJ...510..485A}. The quadrupolar event studied by \citet{2012ApJ...750...12S} illustrates that this transfer can have different effects within the same magnetic system. Reconnection at the null point transferred magnetic flux from the central lobe into the lateral lobes. The resulting reduction in stabilizing flux above the large filament beneath the central lobe enabled it to erupt with a CME, whereas the added flux strengthened the confinement in the lateral lobes. The strongly kinked filament beneath the left lobe rose but subsequently fell back, while an upper magnetic structure resembling a blob erupted outward. The event was therefore interpreted as a partial flux rope eruption caused by reconnection at a site within the flux rope but above the apex of the filament. This reconnection allowed the upper magnetic structure to escape while the filament material remained confined. 
Reconnection in a breakout configuration can also act as negative feedback. In the failed breakout eruption reported by \citet{2022A&A...665A..51S}, reconnection between the two lateral loop systems at low heights produced hot loops moving downward that reinforced the central loop system and expanding loops moving upward that continuously supplied magnetic flux to the large loop system at greater heights. In such a configuration, the additional restraining flux in both systems is expected to strengthen the confinement of the core field within the central lobe, making its eruption more difficult.

External reconnection can suppress an eruption more directly when the rising flux rope reconnects with the overlying or peripheral field. Numerical studies have shown that this process can remove flux from the rope, reduce its twist and upward magnetic force, or even split and destroy the rope \citep{2016ApJ...832..106H,2023ApJ...951L..35C,2023MNRAS.525.5857J}. Observational evidence has become increasingly detailed. \citet{2023ApJ...959...67C} found that early breakout and flare reconnection initiated and accelerated an eruption, whereas subsequent external reconnection formed hot loops at greater heights, fragmented the filament, and led to its confinement. For another failed eruption, \citet{2025ApJ...979..113L} and \citet{2025ApJ...988...14O} reported simultaneous heating of the filament and overlying loops, bidirectional motions resembling plasmoids, newly formed loops, slipping brightenings, and a marked reduction in filament twist. Spectroscopic observations have revealed strong Doppler shifts and broad line profiles where an erupting filament reconnects with coronal loops \citep{2025A&A...698A.301J}. Combined Solar Orbiter and Solar Dynamics Observatory (SDO; \citealt{2012SoPh..275....3P}) observations and a magnetohydrodynamic (MHD) simulation constrained by data recently showed that a flux rope driven by kink instability remained in the torus-stable domain and was decelerated as external reconnection with the overlying loops eroded its eruptive magnetic flux \citep{2026ApJ..1006...81F}. Observations from multiple viewpoints indicate that flare reconnection below a rising prominence and external reconnection with the surrounding field can compete in determining whether the eruption escapes \citep{2026NatAs.tmp..107G}. Such reconnection commonly develops at magnetic interfaces with strong connectivity gradients, including quasi-separatrix layers (QSLs) \citep{1995JGR...10023443P, 1996A&A...308..643D, 2002JGRA..107.1164T, 2009PhPl...16l2101P}. These studies suggest that strong confinement may do more than halt a flux rope. By pressing the rising rope against the overlying loops, strong confinement may promote external reconnection that progressively erodes the erupting structure. This coupling between confinement and external reconnection is central to the event investigated in this paper.

Continuous observations tracing a filament from its formation through activation, strong compression, and eventual confinement remain rare, particularly in the lower solar atmosphere. We investigate a failed filament eruption in NOAA Active Region (AR)~12353 on 2015 May 23 using coordinated chromospheric and coronal observations together with nonlinear force-free field (NLFFF) modeling. Two low-lying filamentary structures merged through internal reconnection to form an elongated filament and reconfigure its supporting flux rope field. Energy release from this internal reconnection likely contributed to the initial rapid rise, with possible kink instability providing an additional driver. A strong, highly asymmetric overlying field prevented the filament from escaping and maintained compression between the rising structure and the overlying loops. The sustained compression provided a key condition for current sheet formation and external magnetic reconnection, which progressively removed magnetic flux from the rope and ultimately disrupted its coherent structure. The external reconnection was accompanied by widespread heating of the active region loop system. Section~2 describes the data, Section~3 presents the observational results, Section~4 presents the magnetic field modeling, Section~5 discusses the physical interpretation, and Section~6 summarizes the conclusions.

\section{Data and Instruments}\label{sec:data}
We combined high-resolution ground-based observations from the 1.6~m Goode Solar Telescope (GST; formerly the New Solar Telescope; \citealt{2010AN....331..620G}) at the Big Bear Solar Observatory (BBSO) with space-based observations from SDO. These coordinated data cover the formation, activation, and failed eruption of a filament in NOAA AR~12353 on 2015 May 23.

The GST Visible Imaging Spectrometer (VIS) sampled the H$\alpha$ line center and the $\pm0.4$ and $\pm0.8$~\AA\ line-wing positions with a bandpass of approximately 0.07~\AA, an image scale of $\sim$0\farcs029~pixel$^{-1}$, and a cadence of approximately 13~s \citep{2010AN....331..636C}. Photospheric TiO images were obtained with the Broadband Filter Imager (BFI) at 7057~\AA\ using a bandpass of $\sim$10~\AA, an image scale of $\sim$0\farcs034~pixel$^{-1}$, and a cadence of approximately 15~s. Both the VIS and BFI data were reconstructed using speckle imaging. The GST observations covered 17:05:08--18:47:16 Universal Time (UT), with several interruptions caused by clouds.

The Atmospheric Imaging Assembly (AIA; \citealt{2012SoPh..275...17L}) on SDO provides full-disk extreme ultraviolet (EUV) images with an image scale of 0\farcs6~pixel$^{-1}$ and a cadence of 12~s. We used the 304, 171, 193, and 335~\AA\ images to examine the filament and coronal loop morphology and kinematics. Images from the six coronal passbands at 94, 131, 171, 193, 211, and 335~\AA\ were used for the differential emission measure (DEM) analysis. The Helioseismic and Magnetic Imager (HMI; \citealt{2012SoPh..275..229S}) provided continuum intensity images and line-of-sight magnetograms at a cadence of 45~s, as well as vector magnetograms at a cadence of 720~s. The HMI data have an image scale of $\sim$0\farcs5~pixel$^{-1}$. The continuum images and line-of-sight magnetograms were used to examine photospheric structures and magnetic evolution, whereas the vector magnetograms provided the lower boundary for the nonlinear force-free field extrapolations.

For the analysis involving multiple instruments, the AIA and HMI observations were differentially rotated to a common reference time of 16:30~UT on 2015 May 23 and registered in the same helioprojective coordinate system. The GST TiO images were coaligned with the HMI continuum images by matching sunspots and pores. The H$\alpha$ images were registered to the TiO images using their common field of view coordinates. This procedure placed the GST chromospheric and photospheric observations in the same coordinate system as the SDO data.

The timing of the associated flare was determined from the 1--8~\AA\ soft X-ray flux measured by the Geostationary Operational Environmental Satellite (GOES) X-ray Sensor (XRS; \citealt{1996SPIE.2812..344H}).

\section{Observations and Results} 
\subsection{Overview of the Source Region}\label{sec:source_region}
Figure~1 presents a multiwavelength overview of the source region of the failed filament eruption in NOAA AR~12353 on 2015 May 23. The HMI line-of-sight magnetogram shows a strong positive polarity concentration in the eastern part of the active region and a strong negative polarity concentration to the west. Several smaller magnetic patches are distributed between and around the main polarities, giving the core region a mixed-polarity configuration (Figure~1(a)). The HMI continuum image shows the corresponding sunspots and pores (Figure~1(b)). The black dotted boxes mark the core source region covered by the high-resolution GST observations.

\begin{figure*}[!t]
\centering
\includegraphics[width=\hsize]{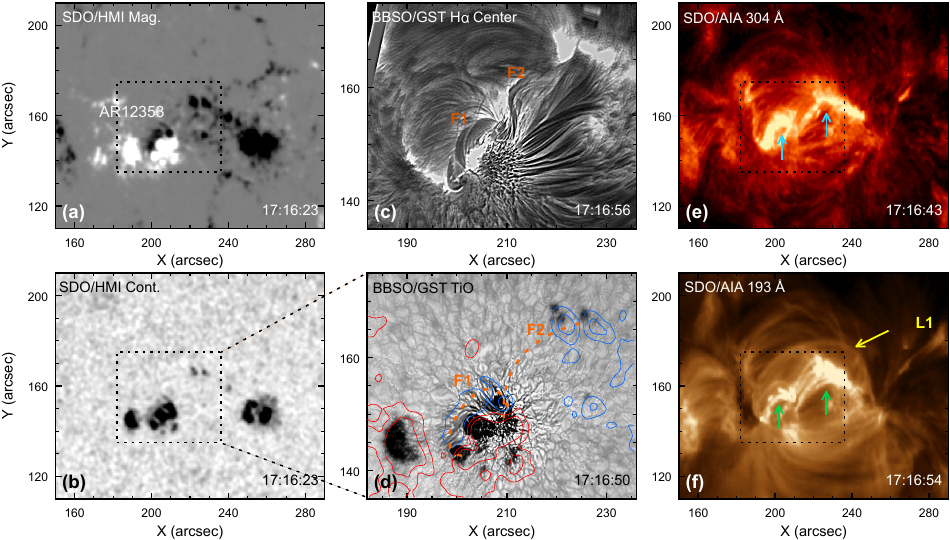}
\caption{Multiwavelength overview of the source region. (a) HMI line-of-sight magnetogram. (b) HMI continuum image. The black dotted boxes in (a), (b), (e), and (f) outline the core region observed by GST. (c) GST H$\alpha$ line-center image showing the filamentary structures F1 and F2. (d) GST TiO image overlaid with HMI line-of-sight magnetic field contours at $\pm200$ and $\pm600$~G. Red and blue contours represent positive and negative magnetic fields, respectively, and the orange dashed curves mark the projected locations of F1 and F2 identified in (c). (e) AIA 304~\AA\ image. The blue arrows mark bright structures above the GST filament region. (f) AIA 193~\AA\ image. The green arrows mark bright coronal loops, and the yellow arrow indicates the larger-scale overlying loop system L1.}
\label{fig:1}
\end{figure*}

The GST H$\alpha$ and TiO images resolve the lower atmospheric structures within this core region (Figures~1(c) and (d)). The H$\alpha$ line-center image shows two low-lying filamentary structures, F1 and F2, whose adjacent ends are close to each other in projection (Figure~1(c)). Figure~1(d) shows the photospheric sunspot and pores in the GST TiO image, together with HMI line-of-sight magnetic field contours. The orange dashed curves mark the projected locations of F1 and F2 identified from the H$\alpha$ image. This comparison shows that their adjacent ends are projected above closely spaced magnetic fields of positive and negative polarity.

The AIA 304 and 193~\AA\ images show the corresponding EUV morphology. Two groups of bright structures are visible in the AIA 304~\AA\ image above the GST filament region, with the more prominent features marked by the blue arrows (Figure~1(e)). The AIA 193~\AA\ image shows bright coronal loops in the core region, as marked by the green arrows (Figure~1(f)). A larger-scale loop system, denoted as L1, lies above the core region. The later observations show that L1 played a major role in confining the rising filament.

\subsection{Formation and Initial Rise of the Filament}
Figure~2 shows the formation and initial rise of filament F. At about 17:30~UT, a localized brightening appeared near the adjacent ends of F1 and F2 in the GST H$\alpha$ $-0.4$~\AA\ image, as marked by the green box in Figure~2(a). F1 and F2 subsequently became connected in this region and formed a longer filament with a twisted appearance, denoted as F (Figure~2(b)). The orientation of the H$\alpha$ fine structures relative to the surrounding magnetic polarities is consistent with dextral chirality, in agreement with the hemispheric chirality rule proposed by \citet{1998SoPh..182..107M}.

\begin{figure*}[!t]
\centering
\begin{interactive}{animation}{Figure2.mp4}

\centering
\includegraphics[width=\hsize]{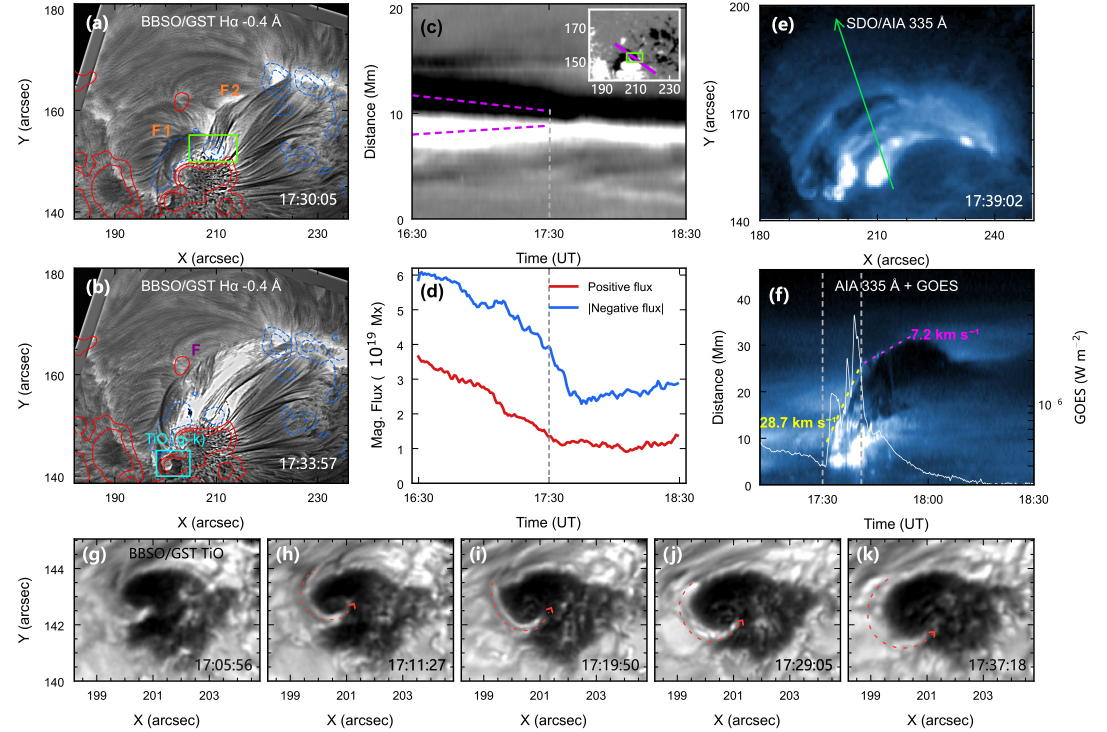}
 \end{interactive}
\global\digitassetfalse
\caption{Formation, magnetic evolution, and initial rise of filament F. (a)--(b) GST H$\alpha$ $-0.4$~\AA\ images showing F1 and F2 merging into F, overlaid with HMI line-of-sight magnetic field contours. Red and blue contours represent positive and negative magnetic fields, respectively. The green box in (a) marks the localized brightening, and the cyan box in (b) outlines the eastern footpoint region enlarged in (g)--(k). (c) HMI time-distance map along the purple slit shown in the inset. The purple dashed lines trace the converging opposite-polarity magnetic features. (d) Positive magnetic flux and the absolute value of the negative magnetic flux within the green box in the inset of (c). (e) AIA 335~\AA\ image of the rising F. (f) Time-distance map along the green arrow in (e), overlaid with the GOES 1--8~\AA\ soft X-ray flux. The colored dashed lines trace the two rising phases, and the vertical dashed lines mark the duration of the C2.3 flare. (g)--(k) Enlarged GST TiO images showing the apparent counterclockwise rotational motion of a pore near the eastern footpoint of F. The accompanying animation shows the evolution corresponding to panels~(g)--(k) and covers approximately 17:03:41--17:48:52~UT.}
\label{fig:2}
\end{figure*}

To examine the associated photospheric magnetic evolution, we constructed an HMI time-distance map along the purple slit shown in the inset of Figure~2(c). Two opposite-polarity magnetic features gradually approached each other before about 17:30~UT, indicating converging motion near the adjacent ends of F1 and F2. During the same interval, both the positive magnetic flux and the absolute value of the negative magnetic flux within the green box decreased (Figure~2(d)). The convergence, the simultaneous decrease in opposite-polarity flux, and the localized H$\alpha$ brightening are consistent with flux cancellation and low-atmospheric internal reconnection between F1 and F2. This reconnection likely facilitated their merging into the longer filament F \citep{1992LNP...399...69M,2001ApJ...552..833M}.

We constructed a time-distance map along the direction of rise indicated by the green arrow in Figure~2(e) to trace the kinematic evolution of F (Figure~2(f)). The tracked feature initially rose with an apparent plane-of-sky speed of about 28.7~km~s$^{-1}$, then entered a much slower rising phase at about 7.2~km~s$^{-1}$, and eventually stalled. The GOES 1--8~\AA\ soft X-ray flux shows that the associated C2.3 flare started at about 17:30~UT, peaked at about 17:39~UT, and ended at about 17:41~UT. The close temporal association among the localized H$\alpha$ brightening, the rapid rise of F, and the soft X-ray enhancement is consistent with a contribution from low-atmospheric reconnection to the initial acceleration of F.

The cyan box in Figure~2(b) marks the eastern footpoint region of the newly formed F, which corresponds to the eastern end of F1 before the merging process. Enlarged GST TiO images show that a small pore in this region underwent an apparent counterclockwise rotation during the formation and initial rise of F (Figures~2(g)--(k)). This evolution is more clearly visible in the accompanying animation. The photospheric rotational motion near the eastern footpoint may have contributed to the injection of magnetic helicity and the buildup of twist in the magnetic field supporting the filament \citep{2003SoPh..216...79B,2009ApJ...697.1529F}.

\subsection{Confinement, External Reconnection, and Flux Rope Erosion}
After its initial rapid rise, F continued toward the overlying loop system L1 as it decelerated and eventually stopped. The separation between the two magnetic systems decreased, and their intervening interface became progressively narrower. The interface brightening, oppositely directed apparent motions, newly brightened loops, and weakening of F described below are consistent with current sheet formation and external reconnection. We infer that this reconnection progressively removed magnetic flux from the flux rope carrying the filament and disrupted its coherent structure.

Figures~3(a)--(d) show the evolution following the initial rise of F. After about 17:43~UT, a faint EUV brightening appeared near the upper boundary of F, where the filament was compressed against the overlying loops. The brightening developed into a narrow, elongated structure, as marked by the yellow arrow in Figure~3(b). Several compact, bright plasma features then exhibited apparent motions in opposite directions along F.

\begin{figure*}[!t]	
\centering
\begin{interactive}{animation}{Figure3.mp4}

\centering
\includegraphics[width=\hsize]{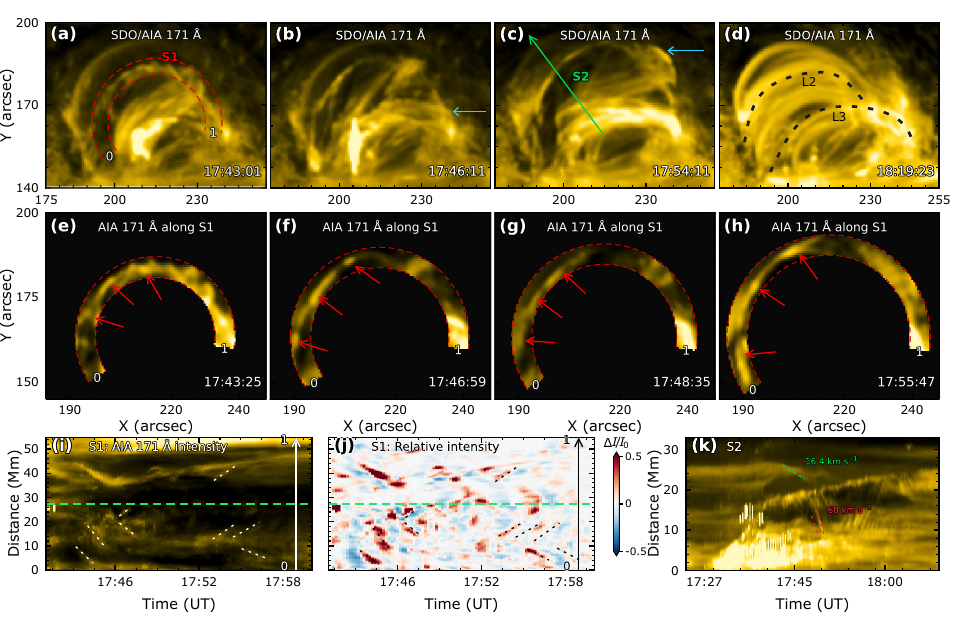}
\end{interactive}
\global\digitassetfalse
\caption{Evolution of the failed filament eruption and associated apparent motions. (a)–(d) AIA 171 Å images. The yellow arrow in (b) marks the elongated brightening near the upper boundary of F, while the cyan arrows in (b) and (c) indicate curved bright structures moving laterally away from the upper part of F on the observer-facing side, accompanied by a modest decrease in projected height. The red dashed curve in (a) and the green arrow in (c) mark the paths S1 and S2, respectively, used for the time–distance analysis. The black dotted curves in (d) outline the newly brightened loop systems L2 and L3. (e)–(h) Straightened AIA 171 Å images along S1, with the red arrows indicating oppositely directed apparent motions. (i) Time–distance map along S1, showing oppositely directed bright tracks fitted by the white dashed lines. (j) Relative-intensity time–distance map along S1, highlighting the same moving features, with the inclined tracks outlined by the black dashed lines. (k) Time–distance map along S2, showing the apparent motions of the bright structures and the overlying loops. The red and green dashed lines indicate representative projected speeds of the separating bright structures and the contracting overlying loops, respectively. The accompanying animation shows the evolution corresponding to panels~(a)--(h) and covers approximately 17:30:11--18:20:11~UT.}
\label{fig:3}
\end{figure*}

Several curved bright structures separated laterally from the upper part of F while decreasing modestly in projected height, as marked by the cyan arrows in Figures~3(b) and (c). The clearly resolved structures were located on the observer-facing side of F. We interpret them as newly reconnected loops separating from the original flux rope structure. The main body of F weakened and became difficult to identify. Two newly brightened loop systems, labeled L2 and L3, appeared during the later phase, as outlined by the black dotted curves in Figure~3(d). The localized interface brightening, oppositely directed motions, weakening of F, and appearance of L2 and L3 are consistent with external reconnection that changed the magnetic connectivity and progressively removed magnetic flux from F.

To better visualize the apparent plasma motions,
Figures~3(e)--(h) show the AIA 171~\AA\ emission extracted along
the curved slit S1, which is outlined by the red dashed curves in
Figure~3(a). The red arrows indicate the oppositely directed apparent
motions, which are most clearly visible on the eastern side of F,
where the plasma features move toward the two footpoint regions.
In the time--distance map constructed along the same slit, these
motions appear as inclined bright tracks (Figure~3(i)).
To suppress the slowly varying background emission and enhance the
transient intensity variations, Figure~3(j) shows the corresponding
relative-intensity time--distance map. The relative intensity is
defined as $(I-I_0)/I_0$, where $I_0$ is the 3-min running temporal
median calculated independently at each position along S1. The
inclined tracks are therefore more clearly visible in Figure~3(j), as outlined by the black dashed lines.
The dashed lines mark the tracks used for the velocity measurements. The selected tracks were identified by comparison with the compact bright features marked by the red arrows in Figures~3(e)--(h) and their evolution in the animation. In the eastern part of F, these features move in opposite directions toward the eastern and western footpoints of F.
We performed a linear fit to each marked track and averaged the resulting apparent velocities separately for motions toward the western and eastern footpoints of F. Taking motion toward the western footpoint of F as positive and motion toward its eastern footpoint as negative, we obtain mean plane-of-sky apparent velocities of $+63.1$ and $-72.5$~km~s$^{-1}$, respectively. The magnitudes of these mean velocities lie near the lower end of the range of apparent reconnection-outflow speeds reported in previous observations
\citep{2013NatPh...9..489S,2023ApJ...943..156K,
	2024A&A...687A.190H}.
The accompanying animation also shows brightenings near both footpoints close in time to the apparent arrival of the moving bright features. This association is consistent with possible impact heating and provides additional support for a plasma-flow interpretation. Some of the apparent motions of the bright structures may also include contributions from the apparent descent of bright loops and superposed flows in other loops, particularly on the western side of F. Taken together, these observations favor interpreting the identified compact moving bright features as possible reconnection outflows, some with a plasmoid-like appearance.
	
Figure~3(k) presents the time-distance map constructed along the green arrow in Figure~3(c). The red dashed line traces the displacement of the observer-facing bright structures away from the upper part of F along the slit, yielding an apparent plane-of-sky speed of about 60~km~s$^{-1}$. Their motion was mainly lateral, with only a modest decrease in projected height. A slower apparent contraction of the overlying loops is also visible at about 16.4~km~s$^{-1}$, as indicated by the green dashed line. The available observations do not allow us to determine whether corresponding separating structures are present on the opposite side of F. The loop separation signatures occurred as the main body of F weakened and the newly brightened loop systems L2 and L3 developed. Their temporal association with the other reconnection signatures is consistent with progressive magnetic flux transfer from F and disruption of its coherent structure. The accompanying animation shows these processes more clearly.

To examine the evolution of the cooler filament plasma, we analyzed H$\alpha$ line-center images together with a qualitative Doppler proxy, defined as $D=(I_b-I_r)/(I_b+I_r)$, where $I_b$ and $I_r$ are the intensities at the $-0.8$ and $+0.8$~\AA\ line-wing positions, respectively. For an absorption feature, negative and positive values generally indicate blueward and redward line asymmetries, respectively, but should not be interpreted as calibrated line-of-sight velocities. During the early phase, F displayed a twisted morphology and an apparent counterclockwise rotational motion while rising, accompanied by predominantly blueward signals (Figures~4(a), (b), (e), and (f)). During the later phase, the curved dark threads in the upper part of F exhibited apparent clockwise rotation. This motion is most clearly visible during 17:45--17:52~UT in the animation accompanying Figure~4. At this stage, the twisted structure of F became progressively less distinct (Figures~4(c) and (d)). A large-scale redward structure subsequently developed, as outlined by the yellow dashed curve in Figure~4(g), supporting the interpretation of downward motion of cool material. Persistent redward signals near both filament footpoints, marked by the green arrows in Figure~4(h), further indicate continued drainage toward the lower atmosphere during the late phase of the failed eruption.

\begin{figure*}[!t]
\centering
\begin{interactive}{animation}{Figure4.mp4}

\centering
\includegraphics[width=\hsize]{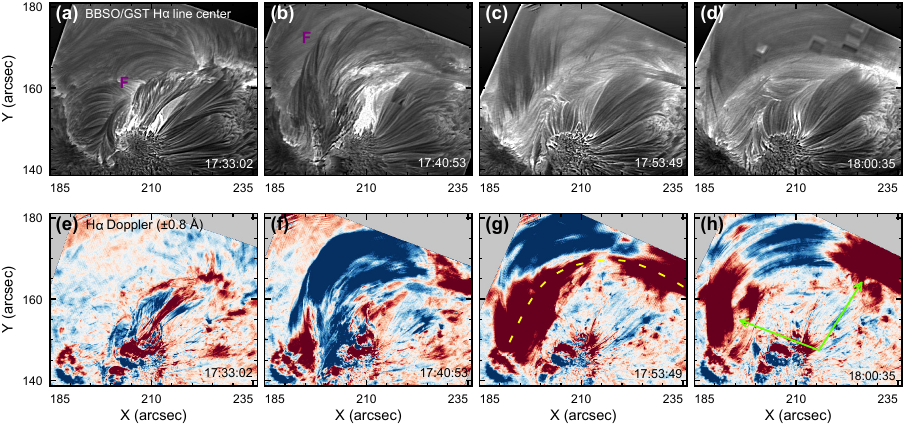}
\end{interactive}
\global\digitassetfalse

\caption{H$\alpha$ evolution of F during the failed eruption. (a)--(d) GST H$\alpha$ line-center images showing the rise, deformation, and progressive weakening of F. (e)--(h) Corresponding qualitative Doppler proxies derived from the intensities at $\pm0.8$~\AA. Blue and red represent blueward and redward line asymmetries, respectively. The yellow dashed curve in (g) outlines a large-scale redward structure, and the green arrows in (h) mark persistent redward signals near the two filament footpoints. The accompanying animation covers approximately 17:30:01--18:06:38~UT.}
\label{fig:4}
\end{figure*}

\subsection{Thermal Evolution and Coronal Heating}
To examine the thermal evolution of the failed eruption and determine whether the morphological signatures of external reconnection were accompanied by localized heating, we performed a differential emission measure (DEM) analysis using the code provided by \citet{2018ApJ...856L..17S}. For optically thin coronal plasma, the intensity observed in AIA passband $i$ is related to the DEM through its temperature response function \citep{2012SoPh..275...41B,2015ApJ...807..143C}.
\begin{equation}
	I_i=\int_{T_{\min}}^{T_{\max}}
	R_i(T)\,\mathrm{DEM}(T)\,dT ,
\end{equation}
where $I_i$ is the observed intensity in passband $i$, $R_i(T)$ is the corresponding temperature response function, and $\mathrm{DEM}(T)$ describes the differential emission measure along the line of sight as a function of temperature. The DEM was reconstructed over the temperature range $\log_{10}(T/\mathrm{K})=5.5$--$7.5$ using the improved sparse inversion method \citep{2015ApJ...807..143C,2018ApJ...856L..17S}. We obtained the emission measure (EM) within each selected temperature interval by integrating the DEM.
\begin{equation}
	\mathrm{EM}=\int_{T_a}^{T_b}\mathrm{DEM}(T)\,dT .
\end{equation}
Because DEM inversion is nonunique, we use the resulting EM maps primarily to compare the relative spatial and temporal evolution of thermal structures rather than to assign a unique temperature to each feature.

Figure~5 presents maps of $\log_{10}[\mathrm{EM}/(\mathrm{cm}^{-5})]$ in three temperature intervals at three representative times. The left, middle, and right columns correspond to 2.5--4.5, 5.0--8.0, and 9.0--20.0~MK, respectively, whereas the top, middle, and bottom rows correspond to 17:25:37, 17:54:01, and 18:11:01~UT, respectively.

\begin{figure*}[!t]
\centering
\begin{interactive}{animation}{Figure5.mp4}

\centering
\includegraphics[width=\hsize]{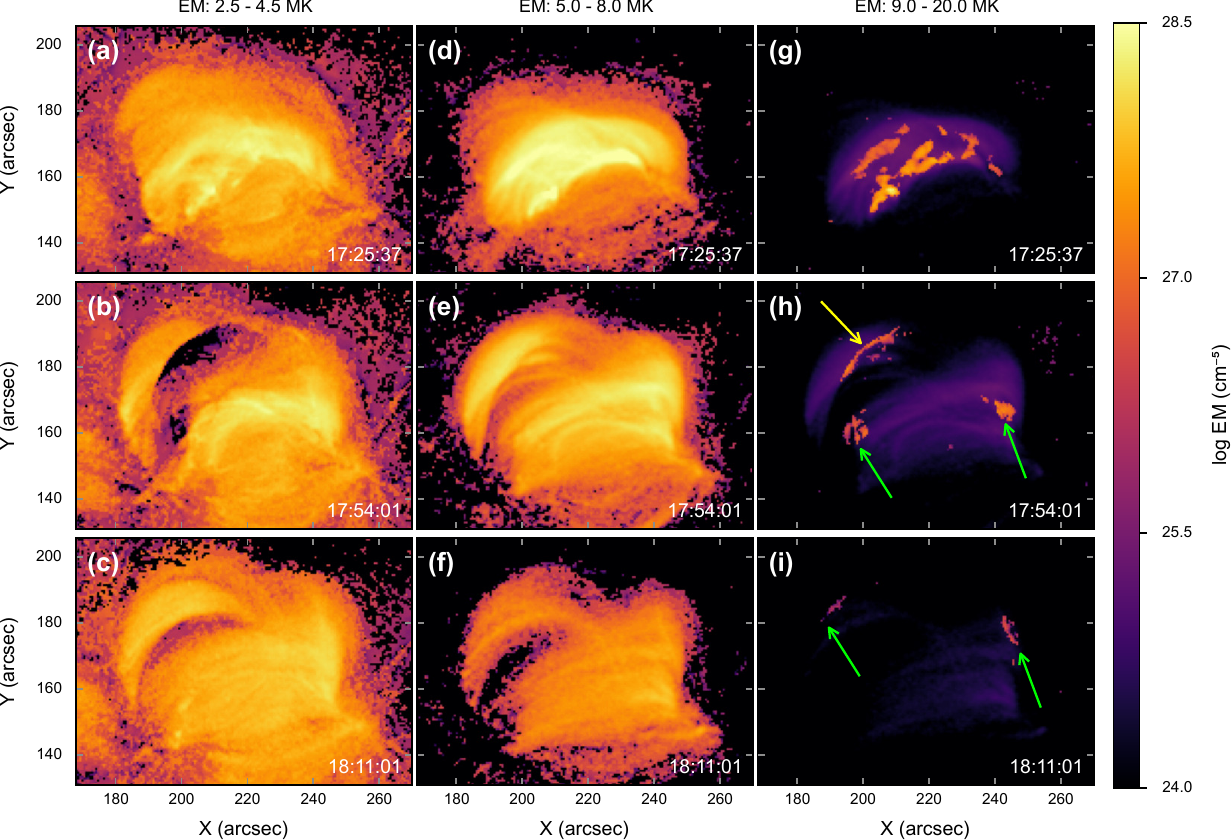}
\end{interactive}
\global\digitassetfalse
\caption{Thermal evolution during the failed eruption. The maps show $\log_{10}[\mathrm{EM}/(\mathrm{cm}^{-5})]$, with the EM calculated by integrating the DEM over 2.5--4.5 MK in (a)--(c), 5.0--8.0~MK in (d)--(f), and 9.0--20.0~MK in (g)--(i). The rows correspond to 17:25:37, 17:54:01, and 18:11:01~UT. The yellow arrow in (h) marks the narrow, unusually hot structure at the F--L1 interface that is consistent with the inferred external reconnection current sheet. The green arrows in (h) and (i) mark compact high-temperature enhancements near the projected footpoints of the loop systems associated with the filament and the overlying or reconnected fields. The accompanying animation covers approximately 17:10:01--18:29:49~UT.}
\label{fig:5}
\end{figure*}

At 17:25:37~UT, before the onset of the external reconnection signatures, the 2.5--4.5~MK EM map outlines the large-scale coronal loop system of the active region (Figure~5(a)). Emission was already present along parts of the preexisting loop system in the 5.0--8.0~MK interval, and the 9.0--20.0~MK map contained patchy high-temperature emission near the filament system before the eruption (Figures~5(d) and (g)). The localized hot patches near the junction of F1 and F2 may reflect energy release during the early phase of the internal reconnection in the low atmosphere that later produced the elongated filament F \citep[e.g.,][]{2015ApJ...804...82C}. They characterize the thermal state before the rapid rise and provide a reference for distinguishing the later, more extended heating and thin structure with high-temperature emission associated with external reconnection.

At 17:54:01~UT, when the signatures of external reconnection were prominent, enhanced 5.0--8.0~MK emission extended over two large loop systems spanning most of the observed active region (Figure~5(e)). This widespread enhancement shows that the heating was not confined to the immediate interaction region between F and L1. A narrow, elongated structure also appeared in the hottest temperature interval considered here, 9.0--20.0~MK, on the eastern side of the main filament body (yellow arrow in Figure~5(h)). It was located in the same general region as the elongated AIA 171~\AA\ bright structure in Figure~3(b). The thin morphology, unusually strong high-temperature emission relative to the surroundings, and spatial association with the other reconnection signatures are consistent with an external reconnection current sheet at this location \citep[e.g.,][]{2017ApJ...835..139S,2018ApJ...853L..15L}.

The green arrows in Figure~5(h) identify additional compact high-temperature enhancements near the projected footpoints of the loop systems associated with the filament and the overlying or reconnected fields. The compact morphology and timing of these sources favor energy deposition by energetic particles accelerated at the reconnection site \citep[e.g.,][]{2011SSRv..159...19F,2011SSRv..159..107H,2012A&A...548A..89N}. Such particles can propagate along newly reconnected field lines and convert their kinetic energy into heat in the denser plasma near the loop footpoints. Thermal conduction along the magnetic field from the hot reconnection region may also contribute, but it would distribute thermal energy along the reconnected loops and may therefore be more relevant to the widespread loop system heating than to the compact footpoint sources alone. Because no direct diagnostic of nonthermal particles is available for this event, the interpretation based on particle precipitation remains inferential.

By 18:11:01~UT, the two large loop systems remained visible in the 5.0--8.0~MK EM map (Figure~5(f)), indicating that the active region loop heating persisted beyond the strongest interface activity. In contrast, the narrow enhancement associated with the current sheet in the 9.0--20.0~MK interval had largely faded, demonstrating its transient nature (Figure~5(i)). Compact high-temperature emission remained detectable near the projected footpoint regions marked by the green arrows. Taken together with the interface brightening, apparent bidirectional motions, and newly formed loop systems in Figure~3, this temporal and spatial evolution supports heating of the active region loop system by external reconnection. It also distinguishes the transient and unusually hot current sheet feature from the more extended and longer-lived loop and footpoint heating produced by the reconnection.

\section{Magnetic Field Modeling}
\subsection{NLFFF Extrapolation and Modeled Magnetic Configuration}
To examine the three-dimensional magnetic structure and topology of the source region, we performed nonlinear force-free field (NLFFF) extrapolations using photospheric vector magnetic fields observed by SDO/HMI. Before the extrapolation, the 180$^\circ$ ambiguity in the transverse field was resolved with the minimum-energy method, which selects the azimuthal solution by minimizing a functional involving the magnetic field divergence and vertical electric current density \citep{1994SoPh..155..235M,2006SoPh..237..267M,2009SoPh..260...83L}. The disambiguated vector field was then transformed into the local heliographic coordinate system using the rotation matrix method of \citet{2017ScChD..60.1408G} to reduce projection effects. The resulting vector magnetograms were used as the lower boundary conditions. The computational domain consisted of a uniform Cartesian grid of $200 \times 80 \times 80$ points with $\Delta x=\Delta y=\Delta z=0.5^{\prime\prime}$.

The extrapolations were performed with the magnetofrictional relaxation method of \citet{2016ApJ...828...82G,2016ApJ...828...83G}. In this method, an artificial velocity, mainly driven by the Lorentz force and controlled by friction, relaxes the magnetic field toward a force-free state. Magnetofrictional modeling has been used to reconstruct successive coronal magnetic configurations during eruptive and confined events \citep{2011ApJ...734...53S,2016ApJ...817...43S,2018ApJ...868...59L,2021ApJ...923..142C}. Magnetofrictional extrapolation is not a time-dependent MHD simulation. Its relaxation iterations are numerical steps rather than a representation of the physical evolution of the corona. The extrapolations at different observing times were also performed independently. We interpret them as a sequence of quasistatic magnetic configurations and use them to examine changes in field geometry and topology, not to infer the eruption dynamics directly \citep[see also][]{2021ApJ...923..142C}.

We extrapolated the field at 17:00, 17:12, 17:24, and 18:00~UT to follow the magnetic structure associated with F. For consistency, each model was analyzed after 30,000 relaxation iterations. Figure~6(a) displays the 17:00~UT lower boundary vector magnetogram. The grayscale image shows the vertical field, the arrows show the horizontal field, and Slits~1 and 2 define the vertical planes used for the twist number ($T_w$) and squashing factor ($Q$) analyses, respectively.

\begin{figure*}[!t]
\centering
\includegraphics[width=\hsize]{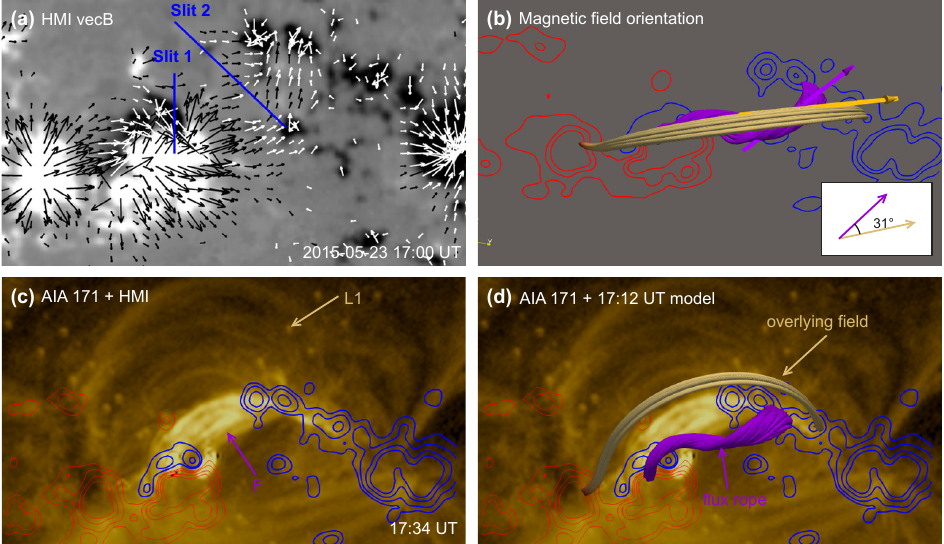}
\caption{Magnetic-field extrapolation results.
	(a) HMI vector magnetogram at 17:00 UT. The grayscale background shows the vertical magnetic field, while the arrows indicate the horizontal magnetic field. The blue lines mark Slits~1--2: Slit~1 is used for the $T_w$ distributions shown in Figures~7(b1)--(b4), Slit~2 for the $Q$ distributions shown in Figures~7(d1)--(d4).
	(b) Three-dimensional configuration of representative flux-rope field lines (purple) and overlying potential-field lines (yellow) from the 17:12 UT model. The inset shows the local angle of approximately $31^\circ$ between their three-dimensional tangent vectors.
	(c) AIA 171~\AA\ image at 17:34 UT, with the magenta and yellow arrows marking the filament F and the overlying loop system L1, respectively.
	(d) The 17:12 UT extrapolated magnetic configuration overlaid on the AIA 171~\AA\ image for comparison with the observed structures.
	The red and blue contours in (b)--(d) mark the positive and negative HMI line-of-sight magnetic fields, respectively, at absolute field strengths of 100, 300, and 500~G.
	} 
\label{fig:6}
\end{figure*}

Figures~6(c) and (d) provide a morphological comparison between the observed filament and loop system and the model. Figure~6(c) shows F and L1 in an AIA 171~\AA\ image at 17:34~UT, together with HMI line-of-sight magnetic field contours. In Figure~6(d), the flux rope field lines from the 17:12~UT NLFFF model (purple) and the corresponding overlying potential field lines (yellow) are plotted on the same 17:34~UT AIA image. The modeled field lines reproduce the broad locations and orientations of F and L1. Because the model and EUV image are separated by 22 minutes, this comparison is intended to establish their large-scale morphological correspondence rather than a simultaneous, one-to-one match.

Figure~6(b) compares the local directions of one representative flux rope field line and one overlying field line in the 17:12~UT model. The angle between their three-dimensional tangent vectors is approximately $31^\circ$. This local, model-dependent value does not represent a global angle between the two flux systems or determine their relative orientation when the external reconnection occurred. The observed rise and rotation of F imply that the local angle may have changed, but its value during the reconnection cannot be determined from the available extrapolations.

\subsection{Evolution of the Flux Rope Twist}
Figures~7(a1)--(a4) show representative field lines tracing the modeled flux rope core at the four extrapolation times. Their changing geometry suggests that the rope became more strongly twisted before the failed eruption and less coherent afterward. We quantified this evolution by calculating the field line twist number, $T_w$, following \citet{2016ApJ...818..148L}.
\begin{equation}
	T_w=\int_L
	\frac{(\nabla\times\mathbf{B})\cdot\mathbf{B}}
	{4\pi B^2}\,dl ,
\end{equation}
where the integration is performed along a magnetic field line. Under the force-free condition, $\nabla\times\mathbf{B}=\alpha\mathbf{B}$,
this expression becomes
\begin{equation}
	T_w=\frac{1}{4\pi}\int_L\alpha\,dl ,
\end{equation}
where $\alpha$ is the force-free parameter \citep[see][]{2006JPhA...39.8321B}. Figures~7(b1)--(b4) show the resulting $T_w$ distributions on the vertical plane defined by Slit~1 in Figure~6(a). The yellow contours denote $T_w=-1$, and the dashed boxes mark the regions selected for quantitative analysis. Within each box, the minimum and mean $T_w$ values were calculated using only pixels with $T_w\leq-1$.

\begin{figure*}[!t]
	\centering
	\includegraphics[width=\hsize]{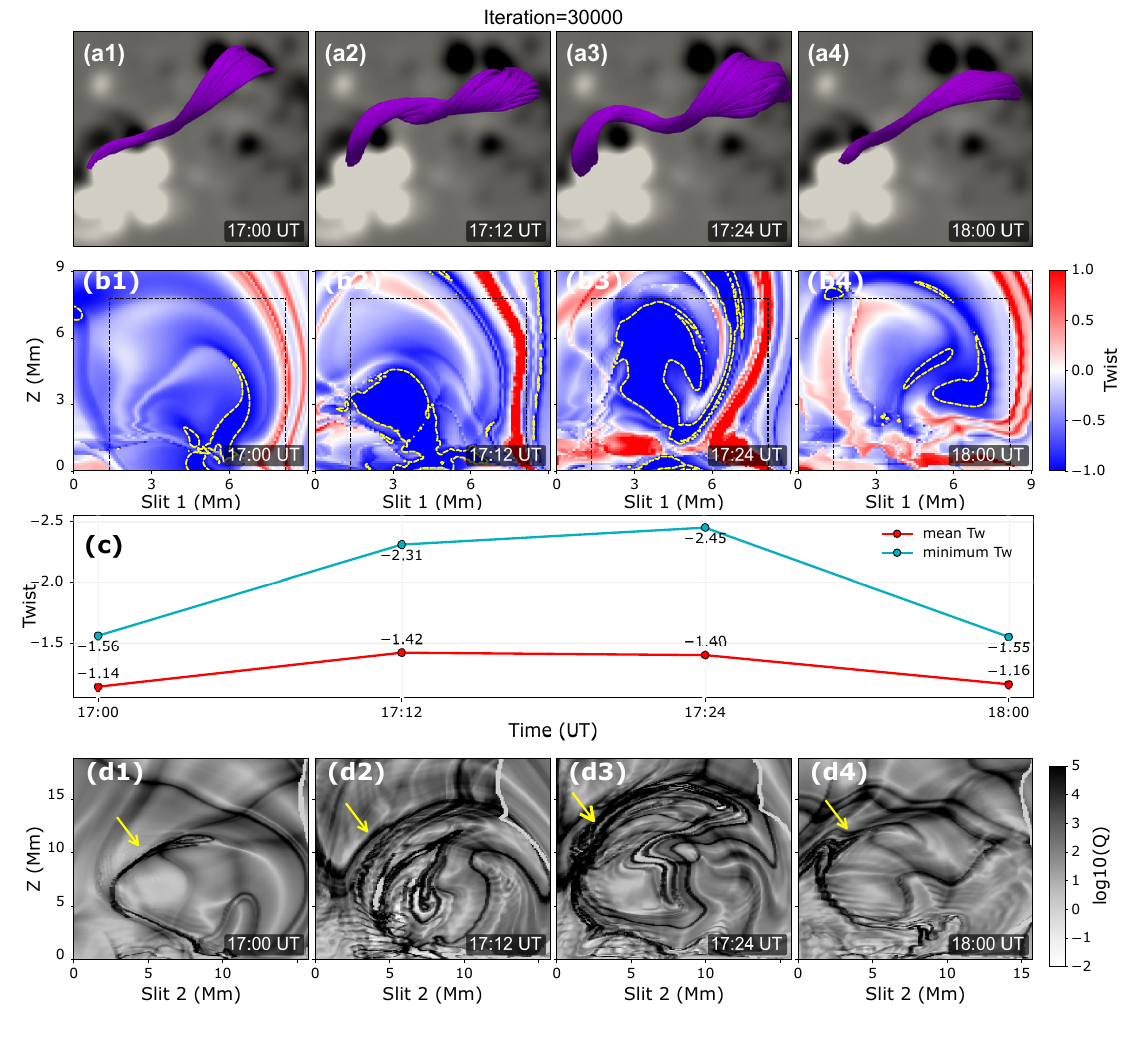}
	\caption{Evolution of the modeled flux-rope structure, magnetic twist, and $Q$ distributions.
		All NLFFF results shown here were obtained after 30,000 relaxation iterations.
		(a1)--(a4) Representative flux-rope field lines from the NLFFF extrapolations at 17:00, 17:12, 17:24, and 18:00 UT, respectively.
		(b1)--(b4) Corresponding distributions of the twist number $T_w$ on the Slit~1 plane. The yellow contours mark $T_w=-1$, and the dashed boxes define the regions used for the $T_w$ statistics shown in (c).
		(c) Temporal evolution of the minimum (cyan) and mean (red) $T_w$ for pixels with $T_w \leq -1$ within the corresponding dashed boxes in (b1)--(b4).
		(d1)--(d4) Corresponding distributions of $\log_{10}Q$ on the Slit~2 plane. The yellow arrows mark the arc-shaped high-$Q$ structure along the upper boundary of the modeled flux rope.
		}
\label{fig:7}
\end{figure*}

Figure~7(c) shows that both twist measures increased in magnitude from 17:00 to 17:12~UT. The most negative $T_w$ continued to change from $-2.31$ at 17:12~UT to $-2.45$ at 17:24~UT, while the mean remained close to $-1.4$. Both measures decreased in magnitude by 18:00~UT. Because these statistics include only pixels with $T_w\leq-1$, they characterize the strongly twisted core rather than the entire flux rope. The magnitude of the mean core twist at 17:12 and 17:24~UT, as well as $|T_w|$ on the most strongly twisted field lines, exceeded the often quoted lower threshold of approximately 1.25 turns ($2.5\pi$) for kink instability in a flux rope with line-tied footpoints \citep[e.g.,][]{2001A&A...367..321B,2004A&A...413L..27T}. The critical twist depends on the rope geometry, line tying, and twist profile \citep{1979SoPh...64..303H,2005ApJ...630L..97T,2016ApJ...832..106H}. The modeled core therefore entered a regime susceptible to kink instability, but this result does not establish kink instability as the eruption trigger.

\subsection{Magnetic Topology and Overlying Field Confinement}
To examine the magnetic connectivity around the modeled flux rope, we calculated the squashing factor, $Q$, which measures the gradient of the field line mapping. High-$Q$ regions identify quasi-separatrix layers (QSLs), across which magnetic connectivity changes rapidly \citep[e.g.,][]{1996A&A...308..643D,2002JGRA..107.1164T,2007ApJ...660..863T}. A high-$Q$ layer does not demonstrate that reconnection has occurred, but it identifies a topologically favorable location for current sheet formation when the field is driven and compressed \citep[e.g.,][]{2006SoPh..238..347A}.

Figures~7(d1)--(d4) show $\log_{10}Q$ on the vertical plane defined by Slit~2 in Figure~6(a). From 17:00 to 17:24~UT, the arc-shaped high-$Q$ layer indicated by the yellow arrows became more pronounced and structured along the upper boundary of the modeled flux rope. By 18:00~UT, this boundary appeared less coherent and more fragmented, consistent with a reorganization of the modeled magnetic connectivity. Because the four NLFFF states were extrapolated independently, these changes should not be interpreted as a direct simulation of the observed compression or reconnection. The model nonetheless places a persistent QSL at the interface where F approached L1. Combined with the observed narrowing of this interface, the later brightening with a current sheet morphology, and the oppositely directed apparent motions, this result supports current sheet formation and external reconnection at this location.

We further used the potential-field model to evaluate the external
magnetic field confining filament F. The decay index is commonly
calculated using the total horizontal potential-field strength,
$B_h=\sqrt{B_x^2+B_y^2}$. When the flux rope rises approximately
vertically and the external horizontal field is predominantly
perpendicular to the rope axis, $B_h$ can approximate the external
field component that restrains the rise. In an actual event, however,
the appropriate field component also depends on the relative
orientations of the external field, the flux-rope axis, and the
eruption path \citep{2021NatCo..12.2734Z}. To account for this geometry
in the present event, we assumed an approximately vertical rise of
the flux rope, along the $z$ direction, based on the observed eruption
direction, and adopted $\overrightarrow{P_0P_1}$ as the reference direction of the
flux-rope axis. We extracted the potential-field component
perpendicular to both this reference axis and the $z$ direction,
denoted its magnitude by $B_\perp$, and calculated its decay index
with height $h$ as $n_\perp=-d\ln B_\perp/d\ln h$,
where $h$ is the height above the photosphere.

Figure~8(a) shows the spatial distribution of $B_\perp$ on a vertical
cross-section. The purple field lines represent the modeled flux rope
at 17:00~UT, and the orange and green contours mark $B_\perp=50$
and $100$~G, respectively. This field component exhibits a pronounced
asymmetry above the two ends of the flux rope. The contours near
the eastern footpoint extend to greater heights, indicating that
$B_\perp$ is stronger on the eastern side than on the western side
at comparable heights in the low corona. Figure~8(b) shows the
distribution of the decay index $n_\perp$ on the same cross-section.
The white and yellow contours mark $n_\perp=1.1$ and $1.5$,
respectively. At greater heights, these contours lie progressively
lower toward the western side, indicating that the corresponding
reference thresholds are reached at lower heights on that side
within this height range. Together with the difference in field
strength between the two sides, these results suggest that the
asymmetric external field may have provided more favorable
conditions for further rise on the western side. This interpretation
is consistent with the observed restricted rise on the eastern side
and more pronounced uplift on the western side (Figures~3 and 4).

\begin{figure*}[!t]
	\centering
	\includegraphics[width=\hsize]{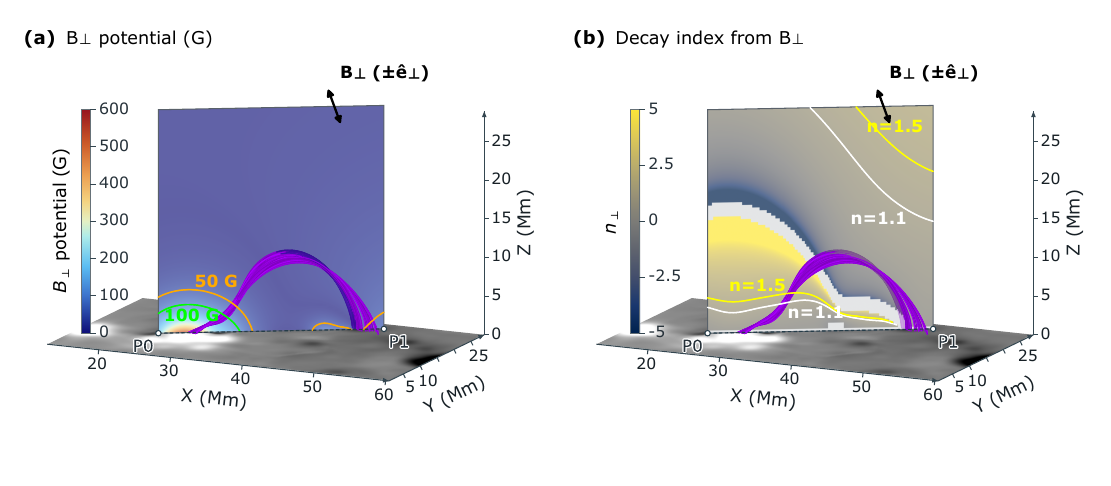}
	\caption{Potential-field diagnostics of magnetic confinement.
		The purple field lines represent the modeled flux rope at 17:00~UT.
		Both panels show the same vertical plane, with $\overrightarrow{P_0P_1}$ defining
		the reference axial direction of the flux rope.
		(a) Magnitude $B_\perp$ of the potential-field component perpendicular
		to both the reference axis and the assumed vertical eruption direction ($z$).
		The black double-headed arrows indicate the projection direction,
		$\pm\hat{\mathbf{e}}_\perp$.
		The orange and green contours mark $B_\perp=50$ and $100$~G, respectively.
		(b) Corresponding decay index,
		$n_\perp=-\mathrm{d}\ln B_\perp/\mathrm{d}\ln h$,
		where $h$ denotes height above the photosphere.
		The white and yellow contours mark $n_\perp=1.1$ and $1.5$, respectively.
		The light-gray bands indicate locations where $n_\perp$ is undefined.} 
\label{fig:8}
\end{figure*}

The decay-index and field-strength distributions reveal a pronounced
east--west asymmetry in the overlying field. At the height of the
modeled flux rope, the low $n_\perp$ together with the substantial
$B_\perp$ indicates strong magnetic confinement. Such confinement
is consistent with the observed deceleration and eventual arrest
of F, in agreement with previous studies showing that strong or
asymmetric overlying fields can favor confined or failed eruptions
\citep[e.g.,][]{2005ApJ...630L..97T,2009ApJ...696L..70L,2016ApJ...832..106H}.
By resisting the upward displacement of the rope, the overlying
field may also have maintained the close contact and compression
between F and L1, thereby helping to establish the thin interface
required for current sheet formation and subsequent external
reconnection.

\section{Discussion}\label{sec:discussion}
The observations and magnetic field modeling reveal two closely coupled roles of the surrounding coronal field in this event. The overlying field confined the rising flux rope and caused the eruption to fail. Its resistance to displacement also maintained a strongly compressed interface between F and L1. We infer that a current sheet formed at this interface and enabled external reconnection to disrupt the rope and heat the active region loop system. Figure~9 summarizes the proposed evolution in four stages.

\begin{figure*}[!t]
\centering
\includegraphics[width=\hsize]{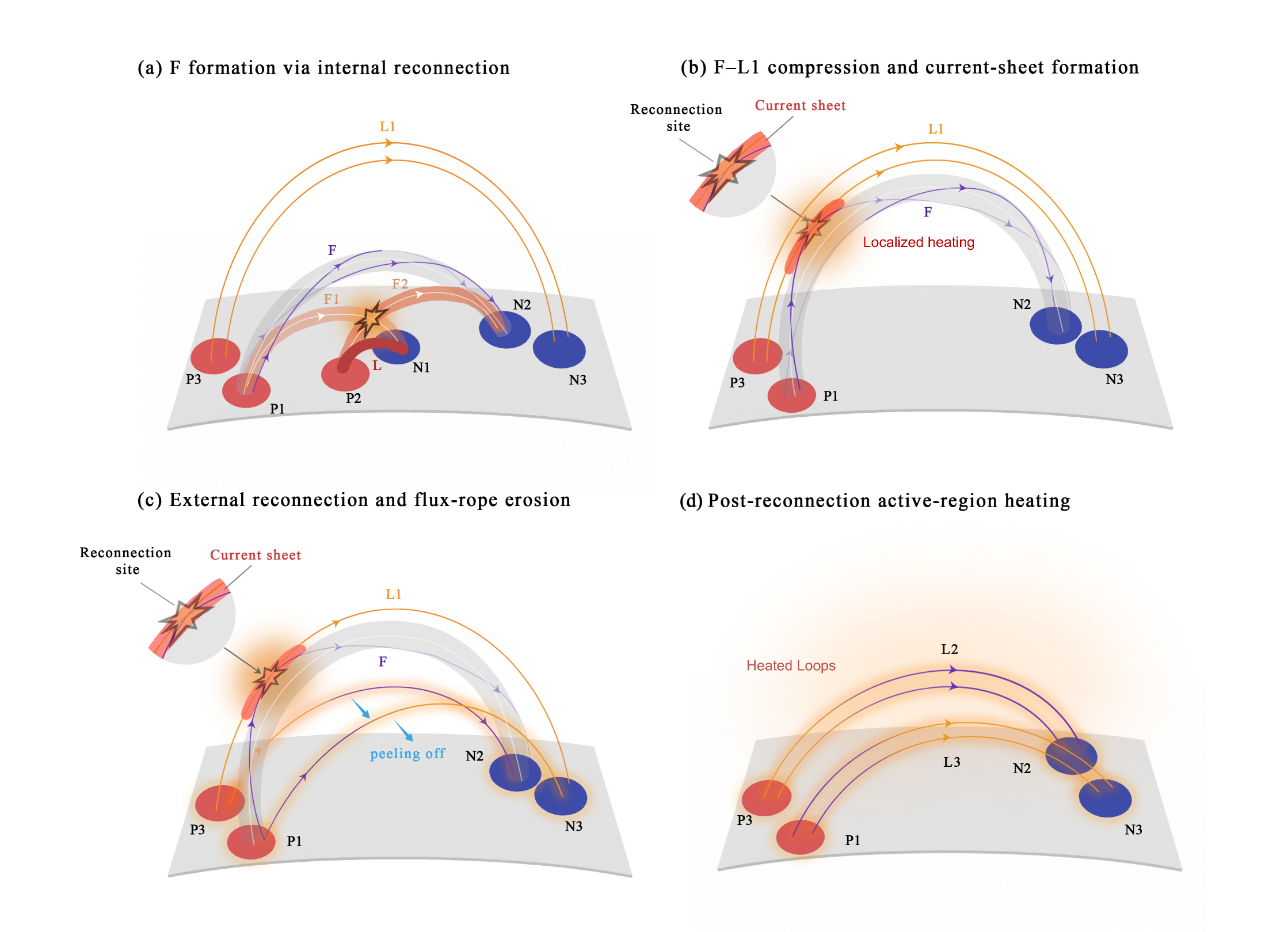}
\caption{Schematic illustration of the proposed magnetic evolution. (a) Internal reconnection between F1 and F2 forms the elongated flux-rope system F. (b) As F rises, it pushes against the overlying loop system L1, compressing the interface between them and forming a localized current sheet. External reconnection begins within the current sheet, accompanied by localized heating. (c) Continued external reconnection exchanges magnetic connectivity between F and L1 and progressively removes magnetic flux from the outer part of F, leading to flux-rope erosion. The blue arrows indicate the peeling-off motion of newly reconnected field lines away from the filament axis. (d) The post-reconnection configuration is dominated by the newly reconnected and heated loop systems L2 and L3, producing the associated active-region heating. Purple and orange field lines represent the magnetic flux systems associated with F and L1, respectively. The translucent gray structure marks the filament axis, while the red elongated structure in panels (b) and (c) represents the current sheet. The yellow cross marks the reconnection site, red and blue circles indicate positive and negative photospheric magnetic polarities, respectively, and orange shading highlights regions of enhanced heating.}
\label{fig:9}
\end{figure*}

\subsection{Flux Rope Formation, Initial Rise, and Confinement}
The convergence and cancellation of opposite-polarity magnetic flux, localized H$\alpha$ brightening, and merging of F1 and F2 support internal reconnection in the low atmosphere as the process that joined the two filamentary structures and reconfigured or strengthened their supporting flux rope field (Figure~9(a)). Similar formation or strengthening of flux ropes through reconnection has been inferred from the tether-cutting scenario and multiwavelength observations \citep{2001ApJ...552..833M,2015ApJ...804...82C,2018ApJ...869...78C,2021ApJ...921L..33Y,2024ApJ...964..125S}. The apparent rotation of the small pore near the eastern footpoint may have injected magnetic helicity and increased the twist of the resulting elongated structure \citep{2003SoPh..216...79B,2009ApJ...697.1529F,2013SoPh..286..453T}.

The rapid initial rise of F was nearly simultaneous with the localized reconnection signatures and the C2.3 flare. Figure~9(b) depicts the later stage in which the rising rope approached and pressed against the overlying loop system. The NLFFF results show that the strongly twisted core entered a regime susceptible to kink instability before the failed eruption. A twist threshold is not universal, and an extrapolated value above its commonly adopted lower limit does not demonstrate that kink instability was triggered \citep{2004A&A...413L..27T,2005ApJ...630L..97T,2016ApJ...818..148L,2016ApJ...832..106H}. The most conservative interpretation is that the internal reconnection in the low atmosphere contributed to the initial acceleration, while the increasing twist and possible kink instability may have provided an additional upward driver.

The subsequent deceleration and arrest of F are consistent with strong confinement by the overlying field, as also illustrated in Figure~9(b). The distributions of $B_\perp$ and $n_\perp$ in Figure~8 show a pronounced east--west asymmetry in the external confining field. At comparable low-coronal heights, $B_\perp$ is stronger near the eastern footpoint, while at greater heights the reference decay-index contours lie lower on the western side. These differences are consistent with the restricted rise on the eastern side and more pronounced uplift on the western side. Although a static extrapolation does not determine the instantaneous Lorentz force on the erupting rope, the field-strength distribution and its variation with height provide diagnostics of the external magnetic confinement \citep{2006PhRvL..96y5002K,2010ApJ...718.1388D,2015ApJ...814..126Z}. The asymmetry may also have contributed to the nonuniform deformation of the rising structure, consistent with previous failed eruptions confined by asymmetric coronal fields \citep{2009ApJ...696L..70L,2018Natur.554..211A}. More generally, recent modeling emphasizes that the three-dimensional strength and geometry of a magnetic cage can constrain an eruption even when a single decay index profile is insufficient \citep{2026A&A...710A.212G}.

\subsection{Role of Confinement in External Reconnection}
The central implication of this event is that the strong overlying field appears to have influenced both the eruption outcome and the development of external reconnection. As F approached L1, the intervening interface narrowed while the rise of F slowed and eventually stopped. Figure~9(b) depicts how the overlying field can maintain this compression and allow the interface to thin into a current sheet. The modeled high-$Q$ layer identifies a persistent, strong connectivity gradient along the upper boundary of the rope, where current sheet formation is expected when neighboring flux systems are driven together \citep{1995JGR...10023443P,1996A&A...308..643D,2002JGRA..107.1164T,2006SoPh..238..347A}. The high-$Q$ layer alone is not evidence of reconnection. Its correspondence with the observed interface brightening, oppositely directed apparent motions, and appearance of L2 and L3 identifies a topologically plausible site for the external reconnection current sheet represented in Figures~9(b) and (c).

Numerical studies have demonstrated several ways in which external reconnection can suppress an eruption. Reconnection with the overlying or surrounding field can remove poloidal or axial flux from a rising rope, reduce its upward magnetic force, separate it from the reconnection region below, or cut and destroy the rope altogether \citep{2016ApJ...832..106H,2023ApJ...951L..35C,2023MNRAS.525.5857J}. The observations and data-constrained MHD simulation reported by \citet{2026ApJ..1006...81F} provide a particularly close comparison. In that event, a rope driven by kink instability remained in the torus-stable domain, was decelerated, and lost eruptive magnetic flux through external reconnection with the overlying loops. Observational studies have also reported filament and loop heating, bidirectional motions resembling plasmoids, new magnetic connections, reduced filament twist, and eventual disappearance or fragmentation of the erupting structure \citep{2016NatPh..12..847L,2023ApJ...959...67C,2025ApJ...979..113L,2025ApJ...988...14O,2025A&A...702A..88T}.

In the present event, the curved bright loop structures moving laterally away from the upper part of F are interpreted as reconnected loops produced as magnetic flux was transferred out of the rope. Only the structures on the observer-facing side are clearly resolved. The available observations do not allow us to determine whether corresponding separating structures are present on the opposite side of F. In the proposed topology, external reconnection changed the original connections P1--N2 and P3--N3 into P1--N3 and P3--N2, producing L2 and L3 (Figures~9(c) and (d)). Here, P and N denote footpoint regions of positive and negative polarity, respectively. The motion of the curved loop structures, the interface brightening, oppositely directed features resembling plasmoids, formation of L2 and L3, fading of F, and decrease in modeled twist after 17:24~UT collectively support progressive flux rope erosion and loss of its coherent magnetic structure. No single signature is conclusive, but their temporal and spatial association provides a consistent external reconnection interpretation. Figure~9(c) shows this intermediate stage of continuing flux transfer. Its two blue arrows illustrate two newly reconnected loops separating from the filament axis.

The $31^{\circ}$ angle measured between representative field line tangents in the 17:12~UT model does not contradict this interpretation. Three-dimensional reconnection across QSLs does not require strictly antiparallel fields and can proceed in the presence of a substantial guide field component \citep{1995JGR...10023443P,2006SoPh..238..347A}. The measured angle is local and predates the observed external reconnection phase. The rise, deformation, and rotation of F could have changed the relative orientation, but the available extrapolations do not determine the angle at the reconnection site and time.

Previous studies have mainly emphasized how external reconnection removes flux from or otherwise weakens an erupting rope. The present observations suggest a condition that may help such reconnection develop. A sufficiently strong overlying field can resist expansion and maintain the compression needed to thin the interface into a current sheet. This proposed sequence is summarized in Figures~9(b) and (c). A weaker or more readily displaced surrounding field might expand with the rope and produce less sustained compression. This inference is consistent with the simulated coupling between overlying field tension and external reconnection \citep{2023ApJ...951L..35C,2023MNRAS.525.5857J,2026ApJ..1006...81F}, but one event cannot establish that strong confinement is either a necessary or sufficient condition. Comparative event studies and time-dependent MHD simulations constrained by data are required to test how current sheet formation depends on the strength, decay rate, and geometry of the surrounding field.

\subsection{Active Region Heating Driven by External Reconnection}
Magnetic reconnection converts magnetic energy into plasma heating, bulk flows, and energy in nonthermal particles that can be transported along the newly reconnected field lines \citep{2011LRSP....8....6S,2011SSRv..159...19F,2011SSRv..159..107H}. In the standard eruptive flare geometry, the dominant reconnecting current sheet forms below the erupting core. We infer that the relevant current sheet in this event formed instead at the external interface between F and L1, allowing the released energy to access the larger loop systems connected to both magnetic structures.

Coronal heating is commonly framed in terms of the sustained energy input required to maintain million-degree plasma \citep{2006SoPh..234...41K, 2012RSPTA.370.3217P}. Proposed mechanisms include numerous small, impulsive releases of magnetic energy driven by photospheric stressing and the dissipation of wave energy \citep{1988ApJ...330..474P, 2008ApJ...677.1348R, 2011ApJ...736....3V}. Observations of braided magnetic structures have provided evidence for small-scale reconnection and nanoflare-scale energy release in warm coronal loops, upper transition region structures, eruptive structures, and superpenumbral fibrils \citep{2018ApJ...854...80H, 2023A&A...679A...9B, 2025ApJ...985...17Z, 2025ApJ...995...94C}. These signatures occur on much smaller spatial and energy scales than the process examined here. The present heating is a transient, event-scale response to external reconnection during a failed eruption rather than evidence that such events maintain the active region corona in a time-averaged sense. Its distinctive feature is the direct coupling among magnetic confinement, current sheet formation, connectivity change, and energy redistribution. The same overlying field that halted the flux rope also sustained the compressed interface where reconnection developed. Energy released at this interface was then transported into two newly connected large-scale loop systems. Without an energy budget or an event occurrence rate, the contribution of this process to the long-term heating requirement of active regions cannot yet be quantified.

The DEM results show two related but spatially distinct thermal responses. Enhanced 5--8~MK emission extended along two large loop systems spanning most of the observed active region, indicating widespread loop heating rather than heating confined to the immediate interaction site. A transient, narrow 9--20~MK feature appeared at the F--L1 interface. Thin structures with high-temperature emission are a recognized observational signature of reconnecting current sheets \citep{2017ApJ...835..139S, 2018ApJ...853L..15L}. In this event, its narrow morphology, high-temperature emission, and correspondence with the interface brightening and oppositely directed motions support its identification as the external reconnection current sheet. The simultaneous heating of an erupting filament and overlying loops reported by \citet{2025ApJ...979..113L} provides a closely related observational example of energy release during external reconnection. Hot arcades or loops have also been observed when activated filament threads reconnect with the ambient field \citep{2023ApJ...958..116W}.

The radiative MHD simulation of \citet{2024NatAs...8..706L} provides a useful comparison with these thermal signatures. In their emerging active region, continuous flux emergence drove repeated reconnection in a current sheet and supplied persistent heating to hot coronal loops. Both studies associate hot loop systems with energy release in a reconnecting current sheet and subsequent energy transport along reconnected fields. The driving and temporal regimes differ. The heating modeled by \citet{2024NatAs...8..706L} was maintained by recurrent reconnection during continued flux emergence and appeared persistent in the time average. The present event records a transient episode driven by compression between a confined rising flux rope and an overlying loop system.

 Compact high-temperature sources near the projected loop footpoints appeared during the phase of external reconnection and persisted after the narrow current sheet feature faded. Their compact morphology and timing favor an interpretation involving energy deposition by reconnection-accelerated particles that propagated along the newly connected field lines and lost their kinetic energy in the denser lower atmosphere \citep{2011SSRv..159...19F, 2011SSRv..159..107H, 2012A&A...548A..89N}. Rapid transition region responses at the footpoints of hot coronal loops have been used to infer energy deposition by nonthermal electron beams in other impulsive heating events \citep{2014Sci...346C.315T}. Thermal conduction along the magnetic field from the hot reconnection region may also have contributed, particularly to the more extended heating along the loop systems. Because no direct hard X-ray, spectroscopic, or other diagnostic of nonthermal particles is available, the particle interpretation remains plausible rather than demonstrated.

External reconnection in this confined event had two linked consequences. It progressively disrupted the flux rope carrying the filament and redistributed the released energy through a large fraction of the active region loop system. Figure~9(d) illustrates the magnetic configuration after reconnection, in which the original coherent flux rope is absent, and two newly connected loop systems appear. The event extends the role of external reconnection beyond determining whether an eruption escapes. External reconnection can also determine where the released energy is deposited and which surrounding coronal structures are heated.

\section{Summary and Conclusions}\label{sec:summary}
The observations and modeling reveal a coupled process in which internal reconnection in the low atmosphere joined two filamentary structures and activated their supporting flux rope field. The strong, asymmetric overlying field halted the rise and maintained compression at the interface between the rope and the surrounding loop system. We infer that a current sheet formed at this interface, enabling external reconnection to remove magnetic flux from the rope and redistribute the released energy throughout the active region loop system. The failed eruption, progressive magnetic disruption, and widespread coronal heating were successive consequences of the interaction between the rising flux rope and the same confining magnetic environment. Four findings support this interpretation.

\begin{enumerate}
    \item \textbf{Filament formation and initial rise.} Convergence, flux cancellation, and localized H$\alpha$ brightening indicate that internal reconnection in the low atmosphere joined two initially separate filamentary structures and reconfigured or strengthened their supporting flux rope field. The reconnection likely contributed to the initial acceleration, while photospheric rotation and the modeled twist suggest that kink instability may also have aided the rise.
    \item \textbf{Confinement and interface compression.} The modeled flux rope remained below the torus instability thresholds examined here and was confined by a strong, asymmetric overlying field. As the rising rope decelerated and stopped, its interface with the overlying loop system became progressively narrower and more strongly compressed, providing the conditions for current sheet formation.
    \item \textbf{External reconnection and flux rope disruption.} The interface brightening, oppositely directed apparent motions, laterally separating reconnected loops, formation of two new loop systems, fading of the original structure carrying the filament, and decrease in modeled twist collectively support external reconnection between the rising flux rope and the overlying field. This reconnection transferred magnetic flux out of the rope and progressively disrupted its coherent structure.
    \item \textbf{Active region heating.} External reconnection produced widespread 5--8~MK emission along the active region loop system and compact high-temperature sources near the projected loop footpoints. A transient and unusually hot 9--20~MK feature at the interface between the flux rope and the overlying field is consistent with the inferred current sheet.
\end{enumerate}

The principal implication is that magnetic confinement can influence not only whether an eruption escapes but also whether a sufficiently compressed interface develops for external reconnection. This coupling between confinement and reconnection provides a testable framework for understanding why some failed eruptions undergo strong magnetic restructuring and widespread coronal heating even though they remain confined.

\begin{acknowledgments}
We gratefully acknowledge the use of observational data from the Solar Dynamics Observatory (SDO) and the Goode Solar Telescope (GST) at the Big Bear Solar Observatory (BBSO). SDO is a mission in the National Aeronautics and Space Administration (NASA) Living With a Star (LWS) program, and we thank the AIA and HMI science teams for providing the data. BBSO operation is supported by U.S. National Science Foundation (NSF) grant AGS-2309939 and the New Jersey Institute of Technology, and GST operation is partly supported by the Korea Astronomy and Space Science Institute and Seoul National University. This work is supported by the National Natural Science Foundation of China (grant Nos. 12573059 and 12173083), the Specialized Research Fund for the State Key Laboratory of Solar Activity and Space Weather, and the Science and Technology Program of Guangdong Province (grant No. 2025B1212050001).
\end{acknowledgments}

\bibliographystyle{aasjournalv7.1}

\end{document}